\documentclass[twocolumn,showpacs, prl]{revtex4}

\usepackage{graphicx}
\usepackage{dcolumn}
\usepackage{bm}

\begin{document}
\title{Spin-valley phase diagram of the two-dimensional metal-insulator transition}

\date{\today}
\author{O.\ Gunawan}
\author{T.\ Gokmen}
\author{K.\ Vakili}
\author{M.\ Padmanabhan}
\author{E.\ P.\ De Poortere}
\author{M.\ Shayegan}

\affiliation{Department of Electrical Engineering, Princeton University, Princeton, NJ 08544}

\begin{abstract}
Using symmetry breaking strain to tune the valley occupation of a two-dimensional (2D) electron
system in an AlAs quantum well, together with an applied in-plane magnetic field to tune the spin
polarization, we independently control the system's valley and spin degrees of freedom and map out
a spin-valley phase diagram for the 2D metal-insulator transition. The insulating phase occurs in
the quadrant where the system is both spin- and valley-polarized. This observation establishes the
equivalent roles of spin and valley degrees of freedom in the 2D metal-insulator transition.
\end{abstract}


\maketitle



\begin{figure*}
\includegraphics[width=160mm]{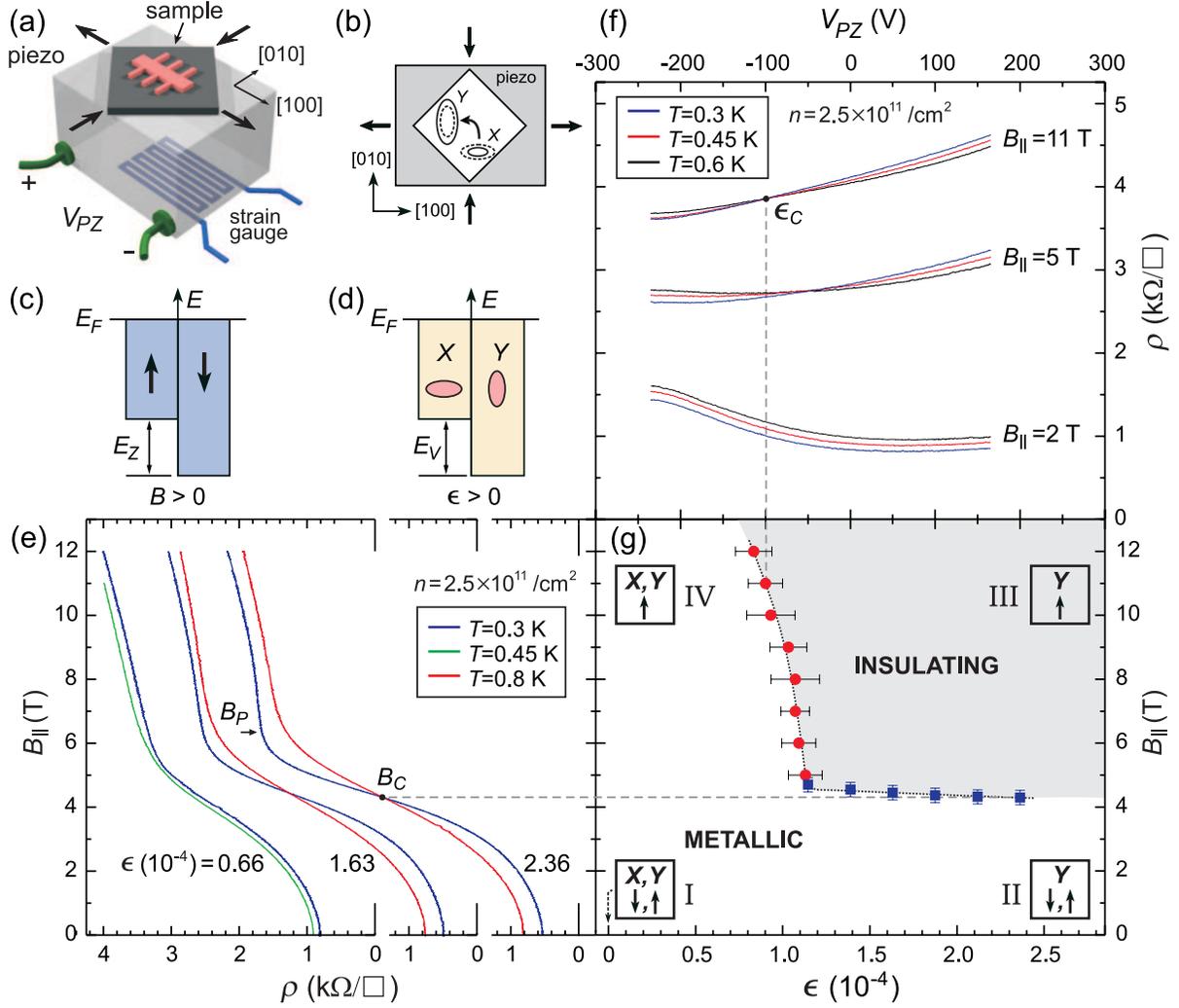} \caption{(color online)
(a)  Schematic of the experimental set-up. The thick arrows indicate the situation where the
sample is under tensile strain along [100] and compressive strain along [010]. (b) The
corresponding electron transfer from the $X$ to the $Y$ valley. (c) and (d) Schematic diagrams
showing the (Zeeman) spin and valley splittings ($E_Z$ and $E_V$) with applied magnetic field or
symmetry-breaking strain, respectively. (e) Magneto-resistance traces at various values of strain.
(f) Piezo-resistance traces at different $B_{\parallel}$. (g) The metal-insulator phase diagram
spanning the four possible combinations of spin and valley occupation, indicated as insets at the
four corners. The dotted arrow at $\epsilon = 0$ ($V_{PZ}=-285$ V) marks the piezo bias at which
the valleys are equally occupied at $B_{\parallel} = 0$. The squares and circles represent the
critical points $B_C$ and $\epsilon_C$, deduced from temperature-dependent traces of (e) and (f),
respectively. The dotted line, delineating the phase boundary, is a guide to the eye only. The
sample is an 11 nm-wide AlAs quantum well with a density of $2.5$$\times$$10^{11}$ cm$^{-2}$ and
mobility (at $\epsilon$ = 0, $B_{\parallel}$ = 0, and $T$ = 0.3 K) of 3.6 m$^2$/Vs.} \label{Fig01}
\end{figure*}

The scaling theory of localization in two dimensions \cite{AbrahamsPRL79}, which predicts an
insulating phase for two-dimensional electron systems (2DESs) with arbitrarily weak disorder, was
challenged by the observation of a metallic temperature dependence ($d\rho/dT > 0$) of the
resistivity, $\rho$, in low-disorder Si metal-oxide-semiconductor field-effect transistors
(Si-MOSFETs) \cite{KravchenkoPRB94}. The associated metal-to-insulator transition (MIT) has
subsequently become the subject of intense interest and controversy \cite{MITRecent}. While
behavior similar to that of Ref.\ \cite{KravchenkoPRB94} has now been reported for a wide variety
of 2D carrier systems such as $n$-AlAs \cite{PapadakisPRB98}, $n$-GaAs \cite{HaneinPRB98},
$n$-Si/SiGe \cite{LaiAPL04PRB05, OkamotoPRB04}, $p$-GaAs \cite{MurzinJETP98, HaneinPRL98,
SimmonsPRL98}, and $p$-Si/SiGe \cite{LamPRB97, ColeridgePRB97}, the origin of the metallic state
and its transition into the insulating phase remain major puzzles in solid state physics.

Several experiments have demonstrated the important role of the spin degree of freedom in the MIT
problem, either in systems with a strong spin-orbit interaction \cite{MurzinJETP98,
PapadakisSci99, YaishPRL00}, or via the application of an external magnetic field to spin polarize
the carriers \cite{SimonianPRL97, OkamotoPRL99, YoonPRL00, PapadakisPRL00, TutucPRL01}. The latter
experiments have shown that a magnetic field applied parallel to the 2DES plane suppresses the
metallic temperature dependence, ultimately driving the 2DES into the insulating regime as the
2DES is spin polarized. The relevance of multiple conduction-band valleys, on the other hand, is
less known. Although it has been discussed theoretically that the occupation of multiple valleys
may also be important \cite{DasSarmaPRB05, PunnooseSci05}, there has been no direct experimental
demonstration. Here we show that the electrons' valley degree of freedom indeed plays a crucial
role, analogous to that of spin. We study a 2DES, confined to an AlAs quantum well, in which we
can independently tune both the spin and valley degrees of freedom. By studying the temperature
dependence of $\rho$ at various degrees of spin and valley polarization, we map out the
metal-insulator phase diagram in this system at a constant density. The 2DES exhibits a metallic
behavior when either the valley or spin are left fully unpolarized, and a minimum amount of {\it
both} spin and valley polarization is required to enter the insulating phase.


We performed experiments on 2DESs confined to modulation-doped, AlAs quantum wells of width 11 nm
and 15 nm \cite{PoortereAPL02}. In these systems, the electrons occupy two conduction-band valleys
of AlAs centered at the edges of the Brillouin zone along the [100] and [010] directions. We
denote these valleys as $X$ and $Y$ [Fig.~\ref{Fig01}(b)], according to the direction of their
major axes. Each valley is characterized by a longitudinal $m_l=1.1 m_0$ and a transverse $m_t
=0.2 m_0$ electron effective mass ($m_0$ is electron mass in vacuum). In our experiments we apply
external strain in the plane of the sample to tune the populations of the $X$ and $Y$ valleys
\cite{ShayeganAPL03, ShkolnikovAPL04, GunawanEsMs06}, and study the evolution of the 2DES as it
becomes valley polarized. This is done by gluing the sample to one side of a stacked piezoelectric
actuator with the sample's [100] crystal orientation aligned with the piezo's poling direction
[Fig.~\ref{Fig01}(a)]. When a voltage bias $V_{PZ}$ is applied to the piezo stack, it expands
(shrinks) along [100] for $V_{PZ}> 0$ ($V_{PZ} < 0$) and shrinks (expands) in the [010] direction,
thus inducing a symmetry breaking strain $\epsilon$ in the sample plane \cite{Epsilondef}. Such
strain lifts the energy degeneracy between the $X$ and $Y$ valleys and electrons are transferred
from one valley to another [Fig.~\ref{Fig01}(b)] while the total density stays constant. To
measure the strain we use a metal-foil strain gauge glued to the piezo's back.

Our measurements were performed in a $^3$He cryostat with a base temperature of 0.3 K. The piezo,
with the glued sample, was mounted in this cryostat on a stage which could be tilted $in$-$situ$
to control the direction of the magnetic field with respect to the sample plane. The stage allows
us to apply a perpendicular magnetic field to characterize the sample via measuring its electron
density and the population of the valleys, and then rotate the sample and apply an in-plane
magnetic field ($B_{\parallel})$ to induce Zeeman spin splitting and eventually spin-polarize the
system. As schematically illustrated in Figs.~\ref{Fig01}(c) and (d), in our experiments
$B_{\parallel}$ and strain play analogous roles as they allow us to tune the spin and valley
polarization of the 2DES, respectively.



The strain vs $B_{\parallel}$ phase diagram of Fig.~\ref{Fig01}(g) captures our main observation.
The origin in this figure, where $\epsilon$ and $B_{\parallel}$ are both equal to zero, represents
the condition where the electrons are distributed equally between the $X$ and $Y$ valleys
\cite{BalancePoint} and have zero spin polarization. The opposite limit, where the 2DES is fully
valley and spin polarized, is reached in the upper right corner of this diagram (quadrant III) for
sufficiently large values of strain and $B_{\parallel}$. In our experiments, we measure the
temperature dependence of $\rho$ as either of the two parameters $\epsilon$ or $B_{\parallel}$ is
kept constant and the other is swept to either spin or valley polarize the 2DES. Examples are
presented in Fig.~\ref{Fig01}(e) where $\rho$ vs $B_{\parallel}$ traces are shown for two
temperatures at three values of strain. For $\epsilon$ equal to 1.63$\times$$10^{-4}$ and
2.36$\times$$10^{-4}$, the traces show that at low $B_{\parallel}$ the 2DES exhibits a metallic
behavior ($d\rho/dT > 0$) while above a critical field ($B_C$) the behavior turns insulating
($d\rho/dT < 0$). These critical fields are marked by blue squares in Fig.~\ref{Fig01}(g). The
traces at $\epsilon$ = 0.66$\times$$10^{-4}$, on the other hand, show that the 2DES remains
metallic in the entire range of applied $B_{\parallel}$. In Fig.~\ref{Fig01}(f), we show traces
where $B_{\parallel}$ is kept fixed while strain is swept continuously. Here we see that at small
values of $B_{\parallel}$ (e.g., 2 T) the sample exhibits a metallic behavior in the entire range
of $V_{PZ}$, while at larger $B_{\parallel}$, the metallic behavior at low $V_{PZ}$ changes to
insulating above a $B_{\parallel}$-dependent critical $\epsilon_C$. These $\epsilon_C$ are
represented by red circles in Fig.~\ref{Fig01}(g). The blue squares and red circles in
Fig.~\ref{Fig01}(g) therefore define the boundary between the metallic and insulating phases of
this 2DES. Figure~\ref{Fig01}(g) qualitatively establishes that the 2DES exhibits a metallic
behavior unless there is a significant amount of spin $and$ valley polarization (quadrant III).

Several features of the data in Figs.~\ref{Fig01}(e), (f), and (g) are noteworthy. The traces in
Fig.~\ref{Fig01}(e), e.g., exhibit a large positive magneto-resistance, implying that, as the 2DES
is made more spin polarized, its resistivity increases. This increase in $\rho$ can be largely
attributed to the loss of screening of the ionized impurities upon spin polarization of the 2DES
\cite{DolgopolovJETP00, HerbutPRB01, DasSarmaPRB05}. We see a more subtle dependence of $\rho$ on
valley polarization [Fig.~\ref{Fig01}(f)]. At small $B_{\parallel}$, when the spins are not
polarized, $\rho$ $decreases$ with increasing valley polarization. This is because we are
measuring $\rho$ along [100] [Fig.~\ref{Fig01}(a)] and transferring electrons into the $Y$ valley
which has a smaller effective mass and therefore larger mobility along [100]. At sufficiently high
$B_{\parallel}$, when the electrons are nearly or fully spin polarized, however, the data of
Fig.~\ref{Fig01}(f) show that $\rho$ increases with strain even though electrons are transferred
into the $Y$ valley. This observation suggests that the screening of impurity potentials becomes
weaker with valley polarization also, consistent with the theoretical calculations
\cite{DasSarmaPRB05}. No matter what the cause of increasing $\rho$, experimentally it is clear
from Figs.~\ref{Fig01}(e) and (f) that, at a fixed low temperature, $\rho$ is largest in the
insulating (III-rd) quadrant of Fig.~\ref{Fig01}(g) where the 2DES is most spin and valley
polarized.

\begin{figure}
\includegraphics[width=85mm]{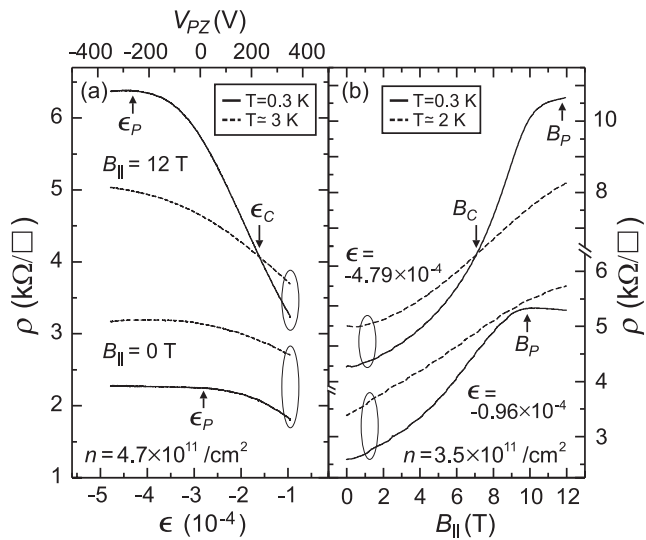} \caption{Data for a 15 nm-wide AlAs quantum well
sample. (a) Piezo-resistance traces at $B_{\parallel}=0$ T, where the 2DES is spin unpolarized,
and $B_{\parallel}=12$ T, where it is (almost) fully spin polarized. (b) Magneto-resistance traces
at $\epsilon=-0.96$$\times$10$^{-4}$, where the 2DES is only slightly valley polarized, and
$\epsilon=-4.79$$\times$10$^{-4}$ where it is fully valley polarized.} \label{Fig02}
\end{figure}

The above observations are further amplified by our measurements on another AlAs sample (15 nm
well-width), presented in Fig.~\ref{Fig02}. Here, in a separate cooldown and by placing the
zero-strain condition at a large positive $V_{PZ}=525$ V, we managed to reach large compressive
strain values ($\epsilon < 0$) along the [100] direction and study the transport properties as the
electrons are transferred into the $X$ valley. Overall, the data are similar to those shown in
Figs.~\ref{Fig01}(e) and (f): When the 2DES is spin or valley degenerate, its metallic behavior
survives as it is made fully valley or spin polarized (bottom sets of traces in Fig.~\ref{Fig02}).
On the other hand, when the 2DES is already significantly spin (or valley) polarized, its metallic
behavior turns insulting as we valley (or spin) polarize it (top sets of traces in
Fig.~\ref{Fig02}) \cite{MITboundary}. Moreover, the piezo-resistance data, shown in
Fig.~\ref{Fig02}(a), show that $\rho$ increases as this charge transfer occurs even at
$B_{\parallel}$ = 0. This is because both the expected loss of screening upon valley polarization,
and the smaller mobility (along the current direction, i.e., [100]) of the $X$ valley to which the
electrons are transferred, lead to a rise in $\rho$. For completeness, in Fig.~\ref{Fig02}(b) we
show the magneto-resistivity of the sample for different values of strain. The remarkable
similarity of the data in Figs.~\ref{Fig02}(a) and (b) attests to the equivalent roles that the
spin and valleys degrees of freedom play in the MIT problem.

The data of Figs.~\ref{Fig01} and \ref{Fig02} reveal yet another important clue regarding the 2D
MIT. In Figs.~\ref{Fig01}(e) and \ref{Fig02}(b), we observe that the positive magneto-resistivity
nearly saturates at high fields, beyond a field $B_P$, as marked in Figs.~\ref{Fig01}(e) and
\ref{Fig02}(b). The field $B_P$ represents the magnetic field beyond which the 2DES is fully
spin-polarized \cite{OkamotoPRL99, YoonPRL00, PapadakisPRL00, TutucPRL01, ShkolnikovPRL04,
DolgopolovJETP00, HerbutPRB01}. The fields $B_C$ and $B_P$ are close to each other but, as pointed
out by Tutuc {\it et al.} \cite{TutucPRL01}, the fact that $B_C < B_P$ implies that the transition
to the insulating phase occurs not at the full spin polarization field but rather when the
population of minority spin electrons reaches below a threshold value. We find that the situation
is similar for the onset of insulating behavior with valley polarization also: As seen in
Fig.~\ref{Fig02}(a), the 2DES turns insulating at a critical strain ($\epsilon_C$) which is
smaller than the strain value at which the system becomes fully valley polarized ($\epsilon_P$)
\cite{epsilon_p}. The magnitude of spin and valley polarization needed to turn the 2DES insulating
appears to depend on the system and density, but it is typically larger than about 50$\%$.

Next, we note that the phase boundary denoted by the red circles in Fig.~\ref{Fig01}(g) is not
vertical but shows a dependence on $B_{\parallel}$. This simply reflects the fact that, thanks to
the finite thickness of the electron layer, the energies of the $X$ and $Y$ valleys slightly shift
with respect to each other as $B_{\parallel}$ increases. The shift comes about because the
parallel field couples to the orbital motion of the electrons \cite{DasSarmaPRL00, TutucPRB03} and
changes the confinement energies of the $X$ and $Y$ valleys. Since $X$ and $Y$ valleys have
anisotropic Fermi contours which are orthogonal to each other, the parallel field, which is
applied along the major axis of one valley ($X$) and perpendicular to the other ($Y$), couples
differently to these valleys, causing a slight shift in their relative energies. We performed
experiments in tilted magnetic fields and, by carefully monitoring the Landau levels for the $X$
and $Y$ valleys as a function of $B_{\parallel}$, measured the shift between the energies of these
valleys \cite{Gokmen07}. Correcting for this shift, the boundary between quadrants III and IV is
nearly vertical.

In summary, a 2DES confined to an AlAs quantum well exhibits a metallic behavior when it is valley
or spin degenerate. The system turns into an insulator when the valley and spins are $both$
sufficiently polarized. The insulating phase has the largest resistivity at a given low
temperature. The data establish experimentally the equivalence of the valley and spin degrees of
freedom in the 2D MIT problem \cite{GunawanEsMs06comment2}. Based on studies of the role of spin,
it has been suggested that temperature dependent scattering and screening are responsible for the
apparent metallic behavior observed in 2D systems at finite temperatures \cite{MurzinJETP98,
PapadakisSci99, PapadakisPRL00, YaishPRL00, TutucPRL01, DasSarmaPRB05}. Our data presented here
provide further credence to such an explanation.

\begin{acknowledgments}
We thank the NSF and ARO for support, and Y.\ P.\ Shkolnikov, E.\ Tutuc, and K.\ Lai for
illuminating discussions.
\end{acknowledgments}


\begin{references}

\bibitem{AbrahamsPRL79}
E. Abrahams, P. W. Anderson, D. C. Licciardello, and T. V. Ramakrishnan, Phys.\ Rev.\ Lett.\ {\bf
42}, 673 (1979).

\bibitem{KravchenkoPRB94}
S. V. Kravchenko, G. V. Kravchenko, J. E. Furneaux, V. M. Pudalov, and M. D'Iorio, Phys.\ Rev.\ B
{\bf 50}, 8039 (1994).

\bibitem{MITRecent}
See for example: E. Abrahams, S. V. Kravchenko, and M. P. Sarachik, Rev.\ Mod.\ Phys.\ {\bf 73},
251 (2001); S. V. Kravchenko and M. P. Sarachik, Rep.\ Prog.\ Phys.\ {\bf 67}, 1 (2004); S. Das
Sarma and E. H. Hwang, Solid State Comm.\ {\bf 135}, 579 (2005).

\bibitem{PapadakisPRB98}
S. J. Papadakis and M. Shayegan, Phys.\ Rev.\ B {\bf 57}, R15 068 (1998).

\bibitem{HaneinPRB98}
Y. Hanein, D. Shahar, J. Yoon, C. C. Li, D. C. Tsui, and H. Shtrikman, Phys.\ Rev.\ B {\bf 58},
R13 338 (1998).

\bibitem{LaiAPL04PRB05}
K. Lai, W. Pan, D. C. Tsui, and Y.-H. Xie, Appl.\ Phys.\ Lett.\ {\bf 84}, 302 (2004); K. Lai, W.
Pan, D. C. Tsui, S. A. Lyon, M. M\"{u}hlberger, and F. Sch\"{a}ffler, Phys.\ Rev.\ B {\bf 72}, R81
313 (2005).

\bibitem{OkamotoPRB04}
T. Okamoto, M. Ooya, K. Hosoya, and S. Kawaji, Phys.\ Rev.\ B {\bf 69}, R41 202 (2004).

\bibitem{HaneinPRL98}
Y. Hanein, U. Meirav, D. Shahar, C. C. Li, D. C. Tsui, and H. Shtrikman, Phys.\ Rev.\ Lett.\ {\bf
80}, 1288 (1998).

\bibitem{SimmonsPRL98}
M. Y. Simmons, A. R. Hamilton, M. Pepper, E. H. Linfield, P. D. Rose, D. A. Ritchie, A. K.
Savchenko, and T. G. Griffiths, Phys.\ Rev.\ Lett.\ {\bf 80}, 1292 (1998).

\bibitem{MurzinJETP98}
S. S. Murzin, S. I. Dorozhkin, G. Landwehr, and A. C. Gossard, JETP Lett.\ {\bf 67}, 113 (1998).

\bibitem{LamPRB97}
J. Lam, M. D'Iorio, D. Brown, and H. Lafontaine, Phys.\ Rev.\ B {\bf 56}, R12 741 (1997).

\bibitem{ColeridgePRB97}
P. T. Coleridge, R. L. Williams, Y. Feng, and P. Zawadzki, Phys.\ Rev.\ B {\bf 56}, R12 764
(1997).

\bibitem{PapadakisSci99}
S. J. Papadakis, E. P. De Poortere, H. C. Manoharan, M. Shayegan, and R. Winkler, Science {\bf
283}, 2056 (1999).

\bibitem{YaishPRL00}
Y. Yaish, O. Prus, E. Buchstab, S. Shapira, G. Ben Yoseph, U. Sivan, and A. Stern, Phys.\ Rev.\
Lett. {\bf 84}, 4954 (2000).

\bibitem{SimonianPRL97}
D. Simonian, S. V. Kravchenko, M. P. Sarachik, and V. M. Pudalov, Phys.\ Rev.\ Lett.\ {\bf 79},
2304 (1997).

\bibitem{OkamotoPRL99}
T. Okamoto, K. Hosoya, S. Kawaji, and A. Yagi, Phys.\ Rev.\ Lett.\ {\bf 82}, 3875 (1999).

\bibitem{YoonPRL00}
J. Yoon, C. C. Li, D. Shahar, D. C. Tsui, and M. Shayegan, Phys.\ Rev.\ Lett.\ {\bf 84}, 4421
(2000).

\bibitem{PapadakisPRL00}
S. J. Papadakis, E. P. De Poortere, M. Shayegan, and R. Winkler, Phys.\ Rev.\ Lett.\ {\bf 84},
5592 (2000).

\bibitem{TutucPRL01}
E. Tutuc, E. P. De Poortere, S. J. Papadakis, and M. Shayegan, Phys.\ Rev.\ Lett.\ {\bf 86}, 2858
(2001).

\bibitem{DasSarmaPRB05}
S. Das Sarma and E. H. Hwang, Phys.\ Rev.\ B {\bf 72}, 205 303 (2005).

\bibitem{PunnooseSci05}
A. Punnoose and A. M. Finkelstein, Science {\bf 310}, 289 (2005).

\bibitem{PoortereAPL02}
E. P. De Poortere, Y. P. Shkolnikov, E. Tutuc, S. J. Papadakis, M. Shayegan, E. Palm, and T.
Murphy, Appl.\ Phys.\ Lett.\ {\bf 80}, 1583 (2002) and references therein.

\bibitem{ShayeganAPL03}
M. Shayegan, K. Karrai, Y. P. Shkolnikov, K. Vakili, E. P. De Poortere, and S. Manus, Appl.\
Phys.\ Lett.\ {\bf 83}, 5235 (2003).

\bibitem{ShkolnikovAPL04}
Y. P. Shkolnikov, K. Vakili, E. P. De Poortere, and M. Shayegan, Appl.\ Phys.\ Lett.\ {\bf 85},
3766 (2004).

\bibitem{GunawanEsMs06}
O. Gunawan, Y. P. Shkolnikov, K. Vakili, T. Gokmen, E. P. De Poortere, and M. Shayegan,
cond-mat/0605692. As detailed in this reference, the valley splitting $E_V$ is equal to the
product of strain and a constant (deformation potential) which can depend on the 2DES density.

\bibitem{Epsilondef}
Here we define strain as $\epsilon=\epsilon_{[100]}-\epsilon_{[010]}$, where $\epsilon_{[100]}$
and $\epsilon_{[010]}$ are the fractional changes in sample's size along the [100] and [010]
directions.

\bibitem{BalancePoint}
Thanks to finite residual stress during the cooling of the sample and the piezo, we need a finite,
cooldown-dependent $V_{PZ}$ to attain the zero-strain condition in our experiments. We determine
this $V_{PZ}$ from our detailed Shubnikov de Haas and coincidence measurements as detailed in Ref.
\cite{GunawanEsMs06}. For the data of Fig.~\ref{Fig01}, we determined the zero-strain condition to
be at $V_{PZ}=-285$ V, and have marked it in Fig.~\ref{Fig01}(g) by a dotted arrow.

\bibitem{DolgopolovJETP00}
V. T. Dolgopolov and A. Gold, JETP Lett.\ {\bf 71}, 27 (2000).

\bibitem{HerbutPRB01}
I. F. Herbut, Phys.\ Rev.\ B {\bf 63}, 113 102 (2001).

\bibitem{MITboundary}
From our measurements in the low-temperature regime (0.3 $< T <$ 0.8 K), we also determined the
boundaries between the metallic and insulating phases for the 15 nm quantum well at a density of
$4.7$$\times$$10^{11}$ cm$^{-2}$: As expected for this higher density, the boundary is pushed to
larger values of $B_{\parallel}$ (10 T) and $|\epsilon |$ (2$\times$$10^{-4}$).

\bibitem{ShkolnikovPRL04}
Y. P. Shkolnikov, K. Vakili, E. P. De Poortere, and M. Shayegan, Phys.\ Rev.\ Lett.\ {\bf 92}, 246
804 (2004).

\bibitem{epsilon_p}
We experimentally determine $\epsilon_P$ from the saturation of the piezo-resistance, as observed
in Fig.~\ref{Fig02}(a), as well as careful measurements of the Shubnikov-de Haas oscillations and
"valley-coincidence" measurements, some of which are described in Ref.~\cite{GunawanEsMs06}.

\bibitem{DasSarmaPRL00} S. Das Sarma and E. H. Hwang, Phys.\ Rev.\ Lett.\ {\bf 84}, 5596 (2000).

\bibitem{TutucPRB03}
E. Tutuc, S. Melinte, E. P. De Poortere, M. Shayegan, and R. Winkler, Phys.\ Rev.\ B {\bf 67},
R241 309 (2003).

\bibitem{Gokmen07}
T. Gokmen, O. Gunawan, Y. Shkolnikov, K. Vakili, E. P. De Poortere, and M. Shayegan, unpublished.

\bibitem{GunawanEsMs06comment2}
The equivalent roles of spin and valley degrees of freedom can also be seen in the equally
enhanced spin and valley susceptibilities in AlAs 2DESs \cite {GunawanEsMs06}.

\end{references}

\end{document}